\documentstyle[12pt,bezier]{article}

\newcommand{\newsection}{    
\setcounter{equation}{0}
\section}
\def\appendix#1{
  \addtocounter{section}{1}
  \setcounter{equation}{0}
  \renewcommand{\thesection}{\Alph{section}}
  \section*{Appendix \thesection\protect\indent \parbox[t]{11.715cm} {#1} }
  \addcontentsline{toc}{section}{Appendix \thesection\ \ \ #1}
  }
\def\eop{\vspace*{\fill}\pagebreak}
\def\tr{\,{\rm tr}\,}
\def\sp{\,{\rm Sp}\,}
\def\e{{\,\rm e}\,}
\def\be{\begin{equation}}
\def\ee{\end{equation}}
\def\bea{\begin{eqnarray}}
\def\eea{\end{eqnarray}}
\def\LA{\left\langle}
\def\RA{\right\rangle}
\newcommand{\rf}[1]{(\ref{#1})}
\newcommand{\eq}[1]{eq.~(\ref{#1})}
\def\cd{{\cal D}}
\def\a{\alpha}
\def\l{\lambda}

\def\eps{\epsilon}
\def\d{\delta}
\def\+{\dagger}
\def\df{\bar{F}}
\def\dw{\bar{W}}
\def\dww{\bar{W}W}
\def\db{B^{\+}}
\def\dbb{B^{\+}B}
\def\G{{\cal G}}

\begin{document}

\begin{flushright}
NBI-HE-96-20  \\
ITEP-TH-15/96 \\
SMI-96-3 \\
cond-mat/9606041\\
June, 1996
\end{flushright}

\begin{center}
\vspace{24pt}
{\large   \bf Supersymmetric matrix models and   branched  polymers}

\vspace{24pt}

{\sl J. Ambj\o rn }

\vspace{6pt}

The Niels Bohr Institute\\
Blegdamsvej 17, DK-2100 Copenhagen \O, Denmark\\

\vspace{24pt}

{\sl Y. Makeenko}\\
\vspace{6pt}
Institute of Theoretical and Experimental Physics,
\\  B. Cheremushkinskaya 25, 117259 Moscow, Russia \\
\vspace{2pt}
and  \\
\vspace{2pt}
The Niels Bohr Institute\\
Blegdamsvej 17, DK-2100 Copenhagen \O, Denmark\\
\vspace{14pt}
and                \\
\vspace{14pt}
{\sl K. Zarembo}     \\

\vspace{6pt}

Steklov Mathematical Institute, \\
Vavilov Street 42, GSP-1, 117966 Moscow, Russia
\\
\vspace{2pt}
and  \\
\vspace{2pt}
Institute of Theoretical and Experimental Physics,
\\  B. Cheremushkinskaya 25, 117259 Moscow, Russia \\
\end{center}

\vspace{18pt}

\vfill

\begin{center}
{\bf Abstract}
\end{center}

\vspace{6pt}

\noindent
We solve a supersymmetric matrix model with a general potential.
While matrix models usually describe surfaces,
supersymmetry enforces a cancellation of bosonic and fermionic
loops and only diagrams corresponding to so-called branched polymers
survive. The eigenvalue distribution of the random matrices near
the critical point is of a new kind.

\vfill
\thispagestyle{empty}
\newpage

\section{Introduction}

The theory of random matrices has been useful in the analysis of
many different physical systems. In particular in condensed-matter
physics one encounters situations where physical quantities may
not depend on the details of a (partly unknown) complicated
Hamiltonian and where the average value of an observable
can be calculated by replacing the average over eigenstates
of the Hamiltonian with an average over an appropriate ensemble of
random matrices. Aspects of vastly different physics
such as electron localisation phenomena
in disordered conductors and semiconductors
\cite{disloc1,disloc2,disloc3,disloc4},
disordered quantum wires \cite{qwires} and integer
quantum Hall effects \cite{qhall} can
be described by random matrix theory. Other applications
can be found in nuclear physics where properties of highly excited
nuclear levels can be described in random matrix language.
Indeed, the very idea of random matrix theory started in nuclear
physics with Wigner's seminal work \cite{wigner} and it was further developed
by Dyson \cite{dyson} and Mehta \cite{mehta}.
For a recent review of the application of random
matrix theory to nuclear physics see \cite{nuclear}.
More recently random matrices have been used in the
study of the low energy chiral properties of QCD \cite{chiral}
and to mention two extremes:
the theory of random matrices has been used in the
spectral theory of quantum mechanical systems with classically chaotic
behaviour \cite{chaos} and in two-dimensional quantum gravity. In the last
case it has even played a derisive role in revealing a whole range of
new non-perturbative phenomena (for recent reviews on this vast topic
we refer to \cite{gravity}).

In many of the applications one is interested in computing ensemble
averages which involve the density of states and correlation functions
of densities of states of the Hamiltonian $H$. The density is
conveniently defined in terms of the trace of the resolvent
of the Hamiltonian:
\be\label{ja1}
G(z) = \frac{1}{N} \tr \frac{1}{z-H}
\ee
where the normalization $N$ indicates that we have already replaced
the real  Hamiltonian with an $N\times N$ random matrix.
The density of states for $H$ is then given by
\be\label{ja2}
\rho(E) = \frac{i}{2\pi} \Big[ G^+ (E) -G^-(E)\Big],
\ee
where  we have introduced the notation
\be\label{ja3}
G^\pm (E) = \frac{1}{N} \tr \frac{1}{E-H \pm i \varepsilon}.
\ee

It is now possible to calculate the  average value over
a number of microscopic energy levels by taking the
average with respect to an appropriate ensemble of random matrices,
typically a Gaussian  ensemble of $N \times N $
Hermitian or orthogonal or complex matrices. For a general
ensemble characterized by a potential $V(H)$, one has
for any observable $f(H)$:
\be\label{ja4}
\LA f(H)\RA = \frac{1}{Z} \int dH \; f(H) \; \exp \Big[-N \tr V(H)\Big],
\ee
where the partition function $Z$ is defined as
\be\label{ja5}
Z = \int dH \; \exp \Big[-N \tr V(H)\Big].
\ee
In \cite{bipz} it was shown how to calculate $G(z)$ in the large $N$ limit
for an arbitrary potential $V(H)$
and the systematic $1/N$ expansion of $G(z)$
was developed in \cite{ackm}. In addition it is known that the large $N$
limit of the connected density--density correlator, or equivalently
\be\label{ja6}
G(z_1,z_2)=
\LA \tr \frac{1}{z_1 -H}\;\tr\frac{1}{z_2 - H} \RA_{\rm conn}
\ee
is {\em  universal} in the sense that it can be expressed as a function
only of the endpoints of the distribution of eigenvalues $\LA\rho (E)\RA$,
independently of the form of the potential $V$. This was first shown
in \cite{ajm} and rediscovered in \cite{zee}.  Essential aspects of the
universality are valid to all orders in the $1/N$ expansion~\cite{ackm}.

In many applications one needs both $G^+(E)$ and $G^-(E)$ and
in these calculations it has been technically convenient to extend
the random matrices to so-called supermatrices, which contain
both Fermionic and Bosonic parts \cite{efetov,ver} (for a review of the
supermatrix theory as used in this context, see
\cite{zuk}). Even in the quantum
gravity applications \cite{enzo,luis}
as well as in purely combinatorial applications
like the Meander problem \cite{MP95} supersymmetric
matrix models have been most useful.

In this paper we will show that a new kind
of supersymmetric random matrix theory
can be used to described  a wide class of so-called branched polymers.

{\it Branched} polymers are generalizations of the conventional {\em linear}
polymers to systems where the polymer chain can branch at each
node. The universal scaling behaviour of branched polymers
is believed to be important for the understanding
of  macromolecules in solutions. Contrary to the linear polymers
the scaling limit is not described by a conformal field theory
\cite{cardy}.
In lattice statistical mechanics they are often modelled
as ``lattice animals'', i.e.\ graphs of connected occupied
sites on a lattice.
Lattice animal exponents in $d$ dimensions have been related
to the Lee-Yang edge singularity  in $d-2$ dimensions
\cite{parisi}. One usually introduces {\it contact interactions} between
occupied neighbour sites which are not linked by a bond and
{\it solvent interactions}
between occupied sites and neighbouring empty sites. In addition
one has a chemical potential associated with the growth of
the branched polymers, i.e.\ the occupation of sites of the connected
graph. One is interested in temperatures where the branched polymer
becomes critical and the number of occupied sites grows to infinity.
The detailed critical behaviour  depends on the contact and solvent
interactions as well as the {\em entropy } of the branched polymers.
Part of the phase diagram can be approximated by taking the
$q \to 1$ limit of an extension of the $q$-states Potts model
\cite{branched}.

The entropy factor is of primary importance for the phase transition.
We use the word {\it entropy} in a generalized sense since it is
intended not only to mean the number of branched polymers of a
given size and shape, but also to take into account that
a given branching at a node will occur with  a definite probability.
In this paper we will concentrate on the analysis of {\em abstract}
polymers, i.e.\ we ignore the embedding on a lattice and show how a
supersymmetric matrix model of random matrices can be used to
determine the universality classes of abstract branched polymers
as a function of the weight attributed to branching at the individual nodes.
This simplification implies that the partition function for
our abstract branched polymer model can
be defined as
\be\label{ja7}
Z = \sum_{BP} w(BP) e^{-\mu |BP|},
\ee
where $|BP|$ denotes the number of sites in the branched polymer
and $\mu$ is the chemical potential. $w(BP)$ is the weight of
branching, i.e.\ it is a product of local weight factors associate
with the branching at each site (or vertex):
\be\label{ja8}
w(BP) = \prod_s g(s).
\ee
We assume that $g(s)$ only depends on the order of the site $s$, i.e.\
the number of neighbours connected to $s$ by links and if $s$ has order
$k$ we write $g_k$ for the weight.

This model can be described by the following supersymmetric random
matrix model:
\be\label{z}
Z=\int\,dW\,d\dw\,\e^{-\frac{1}{\a}N\tr V(\dw W)},
\ee
\be\label{V}
V(x)=x-\sum_{k>1} \frac{g_k}{k} x^k,
\ee
where $W$ and $\dw$ are ``superfields''
\be\label{ja9}
W_a = (B,F),~~~~~~\dw_a = (\db,\df),
\ee
$a=1,2$ and $B$ and $F$ are {\em complex} bosonic and fermionic \sloppy
(i.e.\ Grassmannian) \mbox{$N \times N$} matrices, respectively. Supersymmetry
means rotation between the $B$- and $F$-components \cite{MP95}:
\be\label{susy1}
\d_{\eps}B^{\+}=\df\eps,~~~~\d_{\eps}F=-\eps B,
\ee
\be\label{susy2}
\d_{\bar{\eps}}B=\bar{\eps}F,~~~~\d_{\bar{\eps}}
\df=-\db\bar{\eps},
\ee
where $\eps$ and $\bar{\eps}$ are Grassmann valued {\it matrices} and it
follows trivially that the model is invariant under the supersymmetric
transformation \rf{susy1}--\rf{susy2}. Note that it is a huge symmetry
since $\eps$ is matrix valued.

The rest of this article is organised as follows: in section 2 and 3 we solve
the random matrix model \rf{z} using Schwinger--Dyson equations and
supersymmetry Ward identities and show that the only diagrams
which survive the supersymmetry are so-called cactus diagrams.
In particular, we derive an equation which determines
the possible universality classes. In section 4
we show how the cactus diagrams can be mapped onto
branched polymer graphs and we give a purely combinatorial derivation of the
important eq.~\rf{f1}.
Section 5 discusses various generalizations of the model, while section 6
contains a short discussion of the results obtained.

\newsection{Schwinger--Dyson equations and Ward identities}
\label{1mm}

The supersymmetric correlators of the model defined by the partition
function \rf{z} vanish due to the cancellations between bosonic and
fermionic loops (see eq.~\rf{zero}),
but the correlators of the bosonic matrices are
nontrivial. We shall study the generating function for these correlators:
\begin{equation}\label{loop}
G(\l)=\left\langle\frac{1}{N}\tr\frac{1}{\l-\db B}\right\rangle
=\frac{1}{\l}+\sum_{n=1}^{\infty}\frac{1}{\lambda^{n+1}}G_n
\end{equation}
with
\be
G_n\equiv\left\langle\frac{1}{N}\tr(\dbb)^n\right\rangle .
\label{defGn}
\ee

The Schwinger--Dyson equation for $G(\lambda)$ can be derived from the
equality
\[
\int\,dW\,d\dw\,\frac{1}{N^2}\sum_{ij}
\frac{\partial}{\partial B_{ij}}\left(\e^{-\frac{1}{\a}N\tr V(\dw W)}
B\frac{1}{\lambda-\db B}\right)_{ij}=0
\]
and reads
\begin{equation}\label{sd1}
\left\langle\frac{1}{N}\tr V'(\dw W)\frac{\dbb}{\lambda-\dbb}
\right\rangle=\alpha\lambda\left\langle\left(
\frac{1}{N}\tr\frac{1}{\lambda-\dbb}\right)^2\right\rangle.
\end{equation}

To close this equation, we use the Ward identity which follows from the
invariance of both the action and the measure in \rf{z} under the
supersymmetry transformations \rf{susy1}--\rf{susy2}:
\[
\left\langle\delta_{\eps}\left(F
\frac{1}{\nu-\dww}\frac{1}{\lambda-\dbb}\db\right)\right\rangle=0,
\]
which, after a proper contraction of matrix indices, can be rewritten as
\begin{eqnarray}\label{ward1}
\lefteqn{\left\langle\frac{1}{N}\tr\frac{\dww}{\nu-\dww}
\frac{1}{\lambda-\dbb}\left(1+\frac{1}{N}\tr\frac{\dbb}{\lambda-\dbb}
\right)\right\rangle}\nonumber\\*&=&
\left\langle\frac{1}{N}\tr\frac{1}{\nu-\dww}\frac{\dbb}{\lambda-\dbb}
\,\frac{1}{N}\tr\frac{\dbb}{\lambda-\dbb}\right\rangle.
\end{eqnarray}

This equation can be used to give a formal proof of the fact that the
 partition function is equal to unity and all supersymmetric  correlators
 vanish. Taking the limit $\lambda\rightarrow\infty$ and retaining
 $O\left(1/\lambda\right)$ term in eq.~\rf{ward1}, one finds that
\be\label{zero}
\left\langle\frac{1}{N}\tr\left(\dww\right)^n\right\rangle=0
\ee
for all $n\geq1$.
Since such correlators can be obtained by differentiating  the partition
 function with respect to the coupling constants $g_k$ of the
potential~\rf{V}, the partition function is independent of the potential.

 On the other hand, the correlator entering the left hand side of
eq.~\rf{sd1} can be extracted from eq.~\rf{ward1} in the large $N$ limit,
when factorisation holds. It can be seen expanding eq.~\rf{ward1} in the
powers of $1/\nu$. It is convenient to introduce the function
 \begin{equation}\label{q}
 Q(\nu,\lambda)=\left\langle\frac{1}{N}\tr
 \frac{1}{\nu-\dww}\frac{1}{\lambda-\dbb}\right\rangle
 \end{equation}
 and rewrite eqs.~\rf{sd1} and \rf{ward1} in the large $N$ limit as
 \begin{equation}\label{sd}
 \oint\,\frac{d\omega}{2\pi i}\,\oint\,\frac{d\eta}{2\pi i}\,
 \frac{\omega V'(\eta)Q(\eta,\omega)}{\lambda-\omega}
 =\alpha\lambda G^2(\lambda),
 \end{equation}
and
 \begin{equation}\label{ward}
 \oint\,\frac{d\eta}{2\pi i}\,
 \frac{\eta Q(\eta,\lambda)}{\nu-\eta}\left(
 1+\oint\,\frac{d\omega}{2\pi i}\,
 \frac{\omega G(\omega)}{\lambda-\omega}\right)=
 \oint\,\frac{d\xi}{2 \pi i}
 \frac{\xi Q(\nu,\xi)}{\lambda-\xi}
 \,\,\oint\,\frac{d\omega}{2\pi i}\,
 \frac{\omega G(\omega)}{\lambda-\omega},
 \end{equation}
where the contours of integration encircle all singularities of
$Q$ and $G$, but not $\lambda$, $\nu$ and infinity.

 The integrals in eq.~\rf{ward} can be calculated by taking the residues
 at $\lambda$, $\nu$ and $\infty$. The residue at $\infty$ can be
 found since the asymptotic behaviour of $Q(\nu,\lambda)$ and $G(\lambda)$
 follows from the definitions:
\[
Q(\nu,\lambda)=\frac{1}{\nu\lambda}+O\left(\frac{1}{\lambda^2}\right),
\qquad
 Q(\nu,\lambda)=\frac{G(\lambda)}{\nu}+O\left(\frac{1}{\nu^2}\right)
\]
 and
 \[
G(\lambda)=\frac{1}{\lambda}+O\left(\frac{1}{\lambda^2}\right).
\]
After some algebra one finds from \rf{ward}:
 \begin{equation}\label{q1}
 Q(\nu,\lambda)=\frac{\nu\lambda G^2(\lambda)-\lambda G(\lambda)+1}
 {\nu\lambda\left(\nu G(\lambda)-\lambda G(\lambda)+1\right)}.
 \end{equation}
 This formula is to be substituted in eq.~\rf{sd}. Then the integrals over
 $\eta$ and $\omega$ can be done taking the residues at the
 poles of $Q(\eta,\omega)$ as a function of $\eta$ and taking the
 residues at $\omega=\lambda$ and $\omega=\infty$. The result of this
 calculation reads
 \begin{equation}\label{f'}
 \left(\lambda G(\lambda)-1\right)V'\left(\lambda-\frac{1}{G(\lambda)}
 \right)=\alpha\lambda G^2(\lambda).
 \end{equation}

It is now convenient to parametrise $G(\lambda)$ by
 \begin{equation}\label{pf}
G(\lambda)=\frac{1}{\lambda-x(\lambda)},
 \end{equation}
and from \rf{f'}  we obtain the following equation
for $x(\lambda)$:
 \begin{equation}\label{main}
 xV'(x)=\frac{\alpha\lambda}{\lambda-x}.
 \end{equation}
 If $V(x)$ is a polynomial of $n$--th order, eq.~\rf{main} is algebraic of
 $(n+1)$--th order.

\begin{figure}
\unitlength=0.90mm
\linethickness{0.4pt}
\begin{picture}(153.00,49.00)(0,10)
\bezier{124}(20.00,15.00)(6.00,24.00)(15.00,35.00)
\bezier{48}(15.00,35.00)(20.00,38.00)(25.00,35.00)
\bezier{128}(25.00,35.00)(35.00,25.00)(20.00,15.00)
\bezier{104}(50.00,15.00)(37.00,21.00)(45.00,30.00)
\bezier{52}(45.00,30.00)(50.00,34.00)(55.00,30.00)
\bezier{100}(55.00,30.00)(62.00,21.00)(50.00,15.00)
\bezier{108}(75.00,15.00)(63.00,22.00)(74.00,30.00)
\bezier{124}(76.00,30.00)(89.00,22.00)(75.00,15.00)
\bezier{76}(74.00,30.00)(63.00,33.00)(65.00,40.00)
\bezier{16}(65.00,40.00)(66.00,42.00)(68.00,42.00)
\bezier{72}(68.00,42.00)(73.00,43.00)(75.00,30.00)
\bezier{48}(75.00,30.00)(77.00,40.00)(79.00,41.00)
\bezier{72}(76.00,30.00)(87.00,33.00)(86.00,40.00)
\bezier{40}(86.00,40.00)(84.00,44.00)(79.00,41.00)
\bezier{108}(105.00,15.00)(93.00,22.00)(104.00,30.00)
\bezier{116}(105.00,15.00)(118.00,21.00)(106.00,30.00)
\bezier{108}(104.00,30.00)(93.00,38.00)(104.00,45.00)
\bezier{108}(106.00,30.00)(117.00,37.00)(106.00,45.00)
\bezier{76}(104.00,45.00)(93.00,48.00)(95.00,55.00)
\bezier{16}(95.00,55.00)(96.00,57.00)(98.00,57.00)
\bezier{72}(98.00,57.00)(103.00,58.00)(105.00,45.00)
\bezier{48}(105.00,45.00)(107.00,55.00)(109.00,56.00)
\bezier{72}(106.00,45.00)(117.00,48.00)(116.00,55.00)
\bezier{40}(116.00,55.00)(114.00,59.00)(109.00,56.00)
\bezier{100}(135.00,15.00)(123.00,19.00)(128.00,30.00)
\bezier{100}(135.00,15.00)(147.00,19.00)(142.00,30.00)
\bezier{84}(128.00,30.00)(116.00,32.00)(119.00,40.00)
\bezier{28}(119.00,40.00)(120.00,43.00)(124.00,43.00)
\bezier{80}(123.00,43.00)(131.00,43.00)(129.00,31.00)
\bezier{68}(129.00,31.00)(135.00,37.00)(141.00,31.00)
\bezier{76}(142.00,30.00)(153.00,32.00)(151.00,40.00)
\bezier{28}(151.00,40.00)(149.00,43.00)(146.00,43.00)
\bezier{80}(147.00,43.00)(138.00,42.00)(141.00,31.00)
\put(20.00,15.00){\circle*{2.00}}
\put(50.00,15.00){\circle*{2.00}}
\put(75.00,15.00){\circle*{2.00}}
\put(105.00,15.00){\circle*{2.00}}
\put(135.00,15.00){\circle*{2.00}}
\put(35.00,22.00){\makebox(0,0)[cc]{=}}
\put(63.00,22.00){\makebox(0,0)[cc]{+}}
\put(90.00,22.00){\makebox(0,0)[cc]{+}}
\put(118.00,22.00){\makebox(0,0)[cc]{+}}
\put(152.00,22.00){\makebox(0,0)[cc]{+ $\cdots$}}
\put(14.00,22.00){\line(1,1){11.03}}
\put(13.00,24.00){\line(1,1){10.99}}
\put(13.00,27.00){\line(1,1){8.00}}
\put(14.00,31.00){\line(1,1){4.01}}
\put(15.00,20.00){\line(1,1){11.00}}
\put(17.00,19.00){\line(1,1){9.98}}
\put(19.00,18.00){\line(1,1){7.98}}
\put(21.00,17.00){\line(1,1){5.00}}
\end{picture}
\caption {
   Typical Feynman diagrams for $\left\langle\frac{1}{N}\tr
   \dbb\right\rangle$ in the case where the potential $V(x) =
   x-\frac{1}{2} g_2 x^2-\frac{1}{3} g_3 x^3$.}
 \label{cactus}
\end{figure}
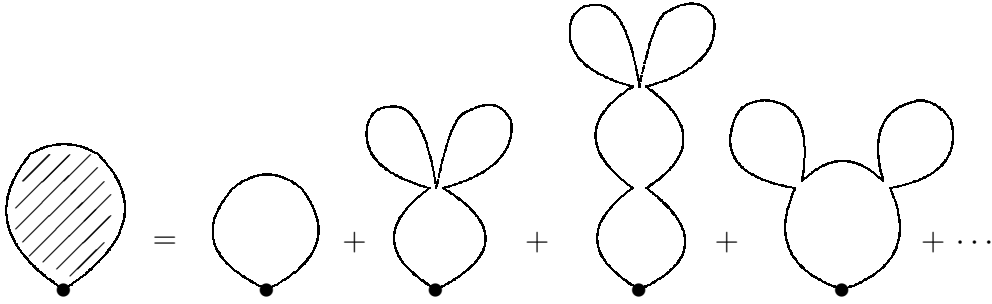

It follows from \rf{pf} and \rf{loop} that
$$
x(\infty)=G_1 ~~\left(\equiv \left\langle\frac{1}{N}\tr\dbb\right\rangle
\right)
$$
so that, taking the limit $\lambda\rightarrow\infty$ in eq.~\rf{main},
we find the closed equation
\begin{equation}\label{f1}
 G_1V'(G_1)=\alpha
\end{equation}
for $G_1$. The surprising fact that we have obtained
a closed equation for the propagator $G_1$ is a
consequence of the cancellations between bosonic and fermionic loops.
In fig.~\ref{cactus} we have shown some of the diagrams which survive
the cancellation. For obvious reasons we name them ``cactus diagrams''.
Note that the diagrams have an orientation: the cactus loops can only
proliferate on the exterior of already existing loops. This is in
contradistinction to related cactus diagrams one encounters in the
large $N$ limit of purely bosonic  or purely fermionic vector models
\cite{onboson,onfermion}.

In addition the Schwinger--Dyson equations for $G_n$ defined by
\eq{defGn}
allow us to express it in terms of $G_k$ with $k<n$.
We give the explicit formulas for the quartic potential
$V(x)=x-\frac{1}{2}gx^2$:
 \begin{equation}\label{f1g}
 G_1=\frac{1-\sqrt{1-4\alpha g}}{2g},
 \end{equation}
 \begin{equation}\label{fng}
 G_n=\frac{2\alpha G_{n-1}+\alpha\sum_{k+l=n-1}G_kG_l+
 g\sum_{s=2}^{n+1}(-1)^s\sum_{k_1+\ldots+k_s=n+1,\,k_i<n}G_{k_1}
 \cdots G_{k_s}}{1-2gG_1}.
 \end{equation}

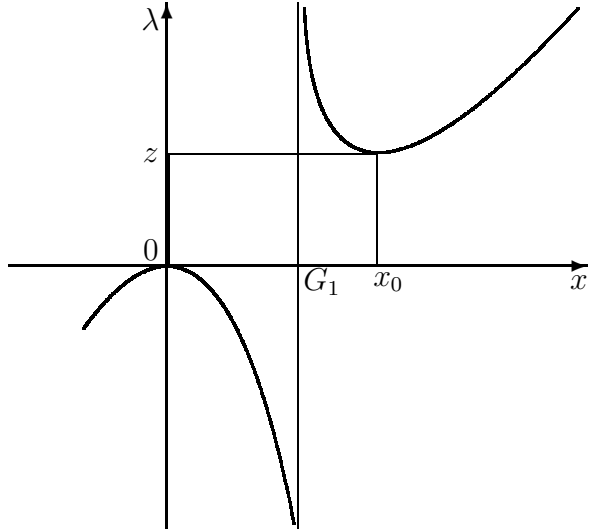
\begin{figure}[tbp]
\unitlength=0.70mm
\linethickness{0.6pt}
\begin{picture}(120.00,108)(-45,15)
\thicklines
\put(10.00,70.00){\vector(1,0){110.00}}
\bezier{476}(24.00,58.00)(50.00,94.00)(64.00,21.00)
\bezier{516}(66.00,119.00)(68.00,64.00)(118.00,119.00)
\put(40.00,20.00){\vector(0,1){100.00}}
\thinlines
\put(65.00,20.00){\line(0,1){100.00}}
\put(40.00,91.28){\line(1,0){40.00}}
\put(80.00,70.00){\line(0,1){21.39}}
\put(39.72,70.00){\rule{0.83\unitlength}{21.39\unitlength}}
\put(85.00,67.00){\makebox(0,0)[rc]{$x_0$}}
\put(73.00,67.00){\makebox(0,0)[rc]{$G_1$}}
\put(37.00,91.00){\makebox(0,0)[cc]{$z$}}
\put(37.00,73.00){\makebox(0,0)[cc]{$0$}}
\put(37.00,117.00){\makebox(0,0)[cc]{$\lambda$}}
\put(120.00,67.00){\makebox(0,0)[rc]{$x$}}
\end{picture}
\caption{
    Function $\lambda(x)$. The interval $(0,z)$ represents an
    eigenvalue support of $\dbb$.} \label{lambda} \end{figure}

 To study the analytical structure of $x(\lambda)$, it is useful to
 consider the inverse function
 \begin{equation}\label{lamb}
 \lambda=\frac{x^2V'(x)}{xV'(x)-\alpha},
 \end{equation}
 which is depicted schematically in fig.~\ref{lambda}.
It has a pole at
 $x=G_1$, as follows from eq.~\rf{f1}, a maximum at zero and a
 minimum at some $x_0>G_1$. Consequently, the function $x(\lambda)$ has
 two branch points at $\lambda=0$ and $\lambda=z\equiv \lambda(x_0)$.
 There may also be  other branch points on the unphysical sheets of
 the Riemann surface of $x(\lambda)$. The branch cut from $0$ to $z$ is to
 be identified with the support of the eigenvalue distribution of the
 positive definite Hermitian matrix $\dbb$, the eigenvalue density being
 given by the discontinuity of $G(\lambda)$ across this cut.

\newsection{Critical behaviour and scaling limit}

Inspection of eq.~\rf{f1} shows that the critical behaviour comes about
 when the equality
 \begin{equation}\label{fc}
 G_cV'(G_c)=\alpha_c
 \end{equation}
 holds simultaneously with
 \begin{equation}\label{dec}
 \left[G_cV'(G_c)\right]'
 =\ldots=
 \left[G_cV'(G_c)\right]^{(m-1)}=0.
 \end{equation}
 Near the critical point $G_1$ behaves as
 \begin{equation}\label{f1c}
 G_1\simeq G_c-\beta^{-1/m}(\alpha_c-\alpha)^{1/m},
 \end{equation}
 where $\beta=(-1)^{m+1}\left[G_cV'(G_c)\right]^{(m)}/m!$. In general
 $m=2$, but the multi-critical points with
 arbitrarily large $m$ can be reached by tuning the potential.
The susceptibility
 $\chi=\frac{\partial G_1}{\partial \alpha}$ at the critical point scales
 as
 \[
 \chi\sim(\alpha_c-\alpha)^{-\gamma}
 \]
 with $\gamma=1-\frac{1}{m}$, which coincides with the (multi)critical
 index of the branched polymers \cite{polymer,multicr}. In the next
 section we will make the mapping on a class of branched polymers explicit.

 Let us turn to the behaviour of the eigenvalue distribution at the
 critical point. Differentiating eq.~\rf{lamb} and expanding around
 $\alpha_c$, one finds that $x_0$, defined by $\lambda'(x_0)=0$
(fig.~\ref{lambda}), scales as $x_0\simeq G_c+(\alpha_c-\alpha)^{1/(m-1)}$.
 Substituting this result into eq.~\rf{lamb}, we obtain that $z\simeq
 G_c\alpha_c(\alpha_c-\alpha)^{-1}$. Thus the eigenvalue distribution
 exhibits a behaviour which is rather unusual for matrix models
 --- at the critical
 point it covers the whole positive real semi-axis. This shows that an
 appropriate variable in the scaling limit is
$$
p=\frac{\lambda(\alpha_c-\alpha)}{\alpha_c G_c},
$$
and that $G(\lambda)$ has the following form near the critical point:
 \begin{equation}\label{fs}
 G(\lambda)\simeq\frac{1}{\lambda}\left[1+\frac{G_c}{\lambda}
 -\frac{(\alpha_c-\alpha)^{1+\frac{1}{m}}}{\alpha_c
 G_c\beta^{\frac{1}{m}}}\,f\left(\frac{\lambda(\alpha_c-\alpha)}
 {\alpha_c G_c}\right)\right].
 \end{equation}
 The coefficient in front of the third term in the square brackets is
chosen to match with eq.~\rf{f1c}.

From \rf{fs} one can extract the scaling
 behaviour of the multi-point correlators of $\dbb$:
 \begin{equation}\label{fns}
 G_n\simeq-\beta^{-\frac{1}{m}}(\alpha_c G_c)^{n-1}
 (\alpha_c-\alpha)^{-n+1+\frac{1}{m}}\,f_n+\delta_{n1}G_c;
 \end{equation}
 and $f(p)$ is a generating function for $f_n$. For the quartic potential
 the scaling behaviour \rf{fns} can be verified directly by using
eq.~\rf{fng}. The generating function $f(p)$ can be found by expansion of
eq.~\rf{f'} or eq.~\rf{main} near the critical point.
A simple calculation leads to
 \begin{equation}\label{fps}
 f(p)=\frac{1}{p}\left(1-\frac{1}{p}\right)^{1/m}.
 \end{equation}

\newsection{Mapping on branched polymers} \label{mobp}

The graphs which survive the supersymmetric cancellation
have an interpretation as branched polymers.
Given a  potential  $V(z)$ with
coupling constants $g_k$ (see \eq{V}) we have seen that
the calculation of  $\LA \dbb \RA$ to leading order in
$1/N$ amounts to a summation
over all graphs of the kind shown in fig.~\ref{cactus}.
The figure can be viewed as a branched polymer in the following
way: $\LA \dbb \RA$ consists of a closed loop of links
with one of the vertices marked. The other vertices are
created by  expanding the interaction
$$\exp \left[-\frac{1}{\alpha} N \tr \Big( V(\dww) -  \dww\Big)\right]$$
in powers of $g_k$ and performing the Gaussian integrals.
If the coupling constant is $g_k$ we attach a ``$k-1$-fold blob''
with weight factor $g_k/\alpha$ if the coupling appears as
$$
\frac{1}{\alpha} \frac{g_k}{k} \; \tr \Bigl( \dww \Bigr)^k.
$$

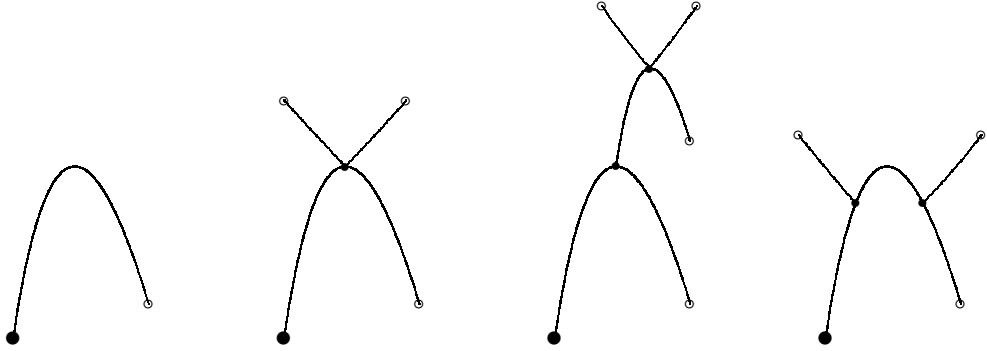
\begin{figure}[tb]
\unitlength=0.90mm
\linethickness{0.4pt}
\begin{picture}(153.87,62.21)(0,10)
\bezier{372}(10.00,10.00)(17.00,58.00)(30.00,15.00)
\bezier{372}(50.00,10.00)(57.00,58.00)(70.00,15.00)
\bezier{372}(90.00,10.00)(97.00,58.00)(110.00,15.00)
\bezier{372}(130.00,10.00)(137.00,58.00)(150.00,15.00)
\bezier{52}(59.02,35.31)(54.36,40.30)(50.03,45.04)
\bezier{52}(59.02,35.31)(63.92,40.38)(68.01,45.04)
\put(68.01,45.04){\circle{1.00}}
\put(50.03,45.04){\circle{1.00}}
\put(69.97,15.03){\circle{1.00}}
\put(59.02,35.31){\circle*{1.00}}
\bezier{204}(99.04,35.36)(102.91,62.21)(109.97,39.10)
\bezier{48}(104.02,49.89)(100.14,54.32)(96.96,59.03)
\bezier{44}(104.02,49.89)(107.62,54.32)(110.94,59.03)
\bezier{52}(134.32,29.97)(130.17,34.81)(126.02,39.93)
\bezier{52}(144.42,29.97)(149.13,34.81)(153.00,39.93)
\put(96.96,59.03){\circle{1.00}}
\put(110.94,59.03){\circle{1.00}}
\put(126.02,39.93){\circle{1.00}}
\put(153.00,39.93){\circle{1.00}}
\put(149.96,15.03){\circle{1.00}}
\put(144.42,29.97){\circle*{1.00}}
\put(134.46,29.97){\circle*{1.00}}
\put(99.04,35.36){\circle*{1.00}}
\put(109.97,15.03){\circle{1.00}}
\put(104.02,49.75){\circle*{1.00}}
\put(109.97,39.10){\circle{1.00}}
\put(30.00,15.00){\circle{1.00}}
\put(10.00,10.00){\circle*{2.00}}
\put(50.00,10.00){\circle*{2.00}}
\put(90.00,10.00){\circle*{2.00}}
\put(130.00,10.00){\circle*{2.00}}
\end{picture}
\caption[cacbp]{The branched polymers corresponding to the cactus
graphs shown in fig.~\ref{cactus}.}
\label{cacbp}
\end{figure}

This procedure can now  be iterated and we get a complete set of cactus
diagrams which is in one-one correspondence with ``chiral''
branched polymers in the following way: the  starting line and the
end line of a cactus loop join at a vertex. Cut the loop open such that
only the stating line is attached to the vertex. This produces a branched
polymer graph which we call ``chiral'' since  branching only occurs
at one side of the open line, corresponding to the fact that
the cactus loops only can be attached to the exterior of already
existing loops.  This is illustrated in fig.~\ref{cacbp}.
We can now write the following combinatorial
identity for the chiral branched  polymers (see fig.~\ref{cbp}):
\be\label{ja51}
G_1 = \alpha + \alpha^2 \frac{1}{\alpha} (-V'(G_1) +1) +
\alpha^3 \frac{1}{\alpha^2} \Big(-V'(G_1)+1\Bigr)^2 +\cdots,
\ee
from which we get eq.~\rf{f1}.
\begin{figure}[tb]
\unitlength=0.90mm
\linethickness{0.4pt}
\begin{picture}(157.00,80.32)(0,30)
\put(10.00,70.00){\line(1,0){5.00}}
\put(21.00,70.00){\line(1,0){5.00}}
\put(26.00,70.00){\line(0,0){0.00}}
\put(40.00,70.00){\line(1,0){15.00}}
\bezier{64}(47.00,70.00)(41.00,79.00)(46.00,80.00)
\bezier{64}(47.00,70.00)(53.00,79.00)(48.00,80.00)
\bezier{8}(47.97,79.99)(47.01,80.10)(46.00,79.99)
\put(70.00,70.00){\line(1,0){15.00}}
\put(100.00,70.00){\line(1,0){15.00}}
\bezier{52}(77.47,70.05)(69.97,75.79)(73.02,78.13)
\bezier{56}(73.02,78.13)(77.36,79.65)(77.47,70.05)
\bezier{48}(77.94,70.05)(79.00,78.95)(82.28,78.13)
\bezier{56}(82.28,78.13)(86.38,76.37)(78.88,70.05)
\bezier{56}(107.00,70.05)(97.81,75.03)(100.52,77.73)
\bezier{60}(100.52,77.73)(103.98,80.32)(106.68,70.05)
\bezier{48}(106.79,70.05)(104.41,79.24)(106.79,79.13)
\bezier{52}(107.00,79.24)(110.24,79.24)(107.22,70.05)
\bezier{56}(107.54,70.05)(110.57,79.89)(114.14,78.05)
\bezier{56}(114.14,78.05)(116.84,75.35)(107.97,70.05)
\bezier{56}(147.00,67.00)(142.00,74.00)(147.00,75.00)
\bezier{56}(147.00,67.00)(152.00,74.00)(147.00,75.00)
\put(33.00,70.00){\makebox(0,0)[cc]{=}}
\put(63.00,70.00){\makebox(0,0)[cc]{$+$}}
\put(93.00,70.00){\makebox(0,0)[cc]{+}}
\put(124.00,70.00){\makebox(0,0)[cc]{$+\cdots =$}}
\put(144.00,70.00){\makebox(0,0)[rc]{{\large$-V'($}}}
\put(151.00,70.00){\makebox(0,0)[lc]{{\large$ )+1$.}}}
\put(20.00,38.00){\makebox(0,0)[cc]{=}}
\put(45.00,38.00){\line(1,0){5.00}}
\put(56.00,38.00){\line(1,0){5.00}}
\put(61.00,38.00){\line(0,0){0.00}}
\put(72.00,38.00){\line(1,0){5.00}}
\put(83.00,38.00){\line(1,0){5.00}}
\put(88.00,38.00){\line(0,0){0.00}}
\put(83.00,38.00){\line(1,0){5.00}}
\put(94.00,38.00){\line(1,0){5.00}}
\put(99.00,38.00){\line(0,0){0.00}}
\put(113.00,38.00){\line(1,0){5.00}}
\put(124.00,38.00){\line(1,0){5.00}}
\put(129.00,38.00){\line(0,0){0.00}}
\put(124.00,38.00){\line(1,0){5.00}}
\put(135.00,38.00){\line(1,0){5.00}}
\put(140.00,38.00){\line(0,0){0.00}}
\put(135.00,38.00){\line(1,0){5.00}}
\put(146.00,38.00){\line(1,0){5.00}}
\put(151.00,38.00){\line(0,0){0.00}}
\put(67.00,38.00){\makebox(0,0)[cc]{$+$}}
\put(106.00,38.00){\makebox(0,0)[cc]{+}}
\put(157.00,38.00){\makebox(0,0)[lc]{$+ \cdots$}}
\put(12.01,47.96){\line(0,0){0.00}}
\put(47.00,70.00){\circle*{0.67}}
\put(77.67,70.00){\circle*{1.33}}
\put(47.00,70.00){\circle*{1.33}}
\put(107.33,70.00){\circle*{1.33}}
\put(12.00,32.99){\circle*{2.11}}
\put(9.67,70.00){\circle*{2.00}}
\put(44.67,38.00){\circle*{2.00}}
\put(72.00,38.00){\circle*{2.00}}
\put(112.67,38.00){\circle*{2.00}}
\put(147.00,66.67){\circle*{2.00}}
\put(40.00,70.00){\circle*{2.00}}
\put(70.00,70.00){\circle*{2.00}}
\put(100.00,70.00){\circle*{2.00}}
\put(55.00,70.00){\circle{1.33}}
\put(85.00,70.00){\circle{1.33}}
\put(115.00,70.00){\circle{1.33}}
\put(61.00,38.00){\circle{1.33}}
\put(99.00,38.00){\circle{1.33}}
\put(151.33,38.00){\circle{1.33}}
\put(26.33,70.00){\circle{1.33}}
\put(15.00,67.00){\framebox(6.00,6.00)[cc]{}}
\put(50.00,35.00){\framebox(6.00,6.00)[cc]{}}
\put(77.00,35.00){\framebox(6.00,6.00)[cc]{}}
\put(88.00,35.00){\framebox(6.00,6.00)[cc]{}}
\put(118.00,35.00){\framebox(6.00,6.00)[cc]{}}
\put(129.00,35.00){\framebox(6.00,6.00)[cc]{}}
\put(140.00,35.00){\framebox(6.00,6.00)[cc]{}}
\put(45.00,79.00){\line(1,0){4.00}}
\put(45.00,77.00){\line(1,0){4.00}}
\put(45.00,75.00){\line(1,0){4.00}}
\put(46.00,73.00){\line(1,0){2.00}}
\put(73.00,77.00){\line(1,0){2.00}}
\put(73.00,75.00){\line(1,0){3.01}}
\put(75.00,73.00){\line(1,0){2.02}}
\put(81.00,77.00){\line(1,0){1.99}}
\put(80.00,75.00){\line(1,0){2.99}}
\put(79.00,73.00){\line(1,0){2.01}}
\put(101.00,77.00){\line(1,0){1.98}}
\put(101.00,75.00){\line(1,0){3.00}}
\put(103.00,73.00){\line(1,0){2.01}}
\put(106.00,78.00){\line(1,0){2.00}}
\put(106.00,76.00){\line(1,0){2.00}}
\put(112.00,77.00){\line(1,0){1.98}}
\put(110.00,75.00){\line(1,0){3.02}}
\put(109.00,73.00){\line(1,0){1.99}}
\put(106.53,74.01){\line(1,0){1.47}}
\put(146.00,74.00){\line(1,0){2.00}}
\put(145.00,72.00){\line(1,0){4.01}}
\put(146.00,70.00){\line(1,0){2.00}}
\bezier{112}(11.96,33.02)(1.33,46.13)(11.96,47.98)
\bezier{116}(12.07,33.12)(23.52,46.95)(12.07,47.98)
\put(9.00,46.00){\line(1,0){6.99}}
\put(8.00,44.00){\line(1,0){9.02}}
\put(8.00,42.00){\line(1,0){9.02}}
\put(9.00,40.00){\line(1,0){6.99}}
\put(10.00,38.00){\line(1,0){4.03}}
\put(11.00,36.00){\line(1,0){2.00}}
\put(25.00,38.00){\line(1,0){8.00}}
\put(25.00,38.00){\circle*{2.00}}
\put(33.00,38.00){\circle{1.33}}
\put(39.00,38.00){\makebox(0,0)[cc]{$+$}}
\end{picture}
\caption[cbp]{The combinatorial equations for the
branched polymer (or cactus) diagrams. The two equations lead
to eq.~\rf{f1}.}
\label{cbp}
\end{figure}

 \newsection{Ising model on branched polymer}

 The purpose of the present section is to study a supersymmetric matrix
 model which describes the Ising model on the branched polymer. It can be
 constructed by introducing a pair of the ``superfields'' $W^1$ and
 $W^2$ whose components transform under supersymmetry
 transformations by \rf{susy1}, \rf{susy2} with the same parameters
 $\eps$ and $\bar{\eps}$ for $W^1$ and $W^2$.
 We choose the interaction potential in the form
 $$
 N\tr\left(\frac{1}{2}g_+\dw_1W^1\dw_1W^1
 +\frac{1}{2}g_-\dw_2W^2\dw_2W^2\right),
 $$ where $g_{\pm}=g\e^{\pm h}$,
 and attach the spin variables to the vertices of the polymer obtained
 from the Feynman diagram for
 $\left\langle\frac{1}{N}\tr\db_1B^1\right\rangle$ as described in
 sec.~\ref{mobp}. The two types of vertices, the ones corresponding to $g_+$
 and $g_-$, are associated with spins
 of the opposite directions.

 The Ising model is recovered
 if the propagators are defined by
 \begin{equation}\label{b+b}
 \left\langle\db_{\alpha ij}B^{\gamma}_{~kl}\right\rangle_{\rm Gauss}
 =\frac{1}{N}\delta_{il}\delta_{jk}D_{\alpha}^{~\gamma}
 \end{equation}
 with
 \begin{equation}\label{dab}
 D_{\alpha}^{~\gamma}=\left(\begin{array}{cc}
 \e^{\beta }&\e^{-\beta }\\ \e^{-\beta }&\e^{\beta }
 \end{array}\right).
 \end{equation}
 The constants $\beta $ and $h$ are to be identified with the inverse
 temperature and an external magnetic field, respectively.

 In what follows
 we shall consider the model without the external field, so that the
 couplings $g_+$ and $g_-$ are equal to each other. The matrix integral
 corresponding to this model has the form
 \begin{equation}\label{z1}
 Z=\int\,\prod_{\alpha=1}^{2}\,dW^{\alpha}\,d\dw_{\alpha}\,
 \e^{-N\tr\left(\dw D^{-1}W-
 \frac{1}{2}g\dw P^+W\dw P^+W-
 \frac{1}{2}g\dw P^-W\dw P^-W\right)},
 \end{equation}
 where
 \begin{equation}\label{p+-}
 P^{\pm}=\frac{1\pm\sigma^3}{2}
 \end{equation}
 and the inverse to~\rf{dab} matrix $D^{-1}$ is
 \begin{equation}\label{dabi}
 \left(D^{-1}\right)^{\gamma}_{~\alpha}=
 \frac{1}{2\sinh{2\beta}}\left(\begin{array}{cc}
 \e^{\beta }&-\e^{-\beta }\\ -\e^{-\beta }&\e^{\beta }
 \end{array}\right)
 \end{equation}
 so that the quadratic part of the exponent in \eq{z1} involves
 \be
 \tr\dw D^{-1}W = \frac{1}{2\sinh{2\beta}} \tr \left[
 \e^{\beta} \left(\dw_1 W^1+\dw_2 W^2 \right)
 -\e^{-\beta} \left(\dw_1 W^2+\dw_2 W^1 \right)
 \right].
 \ee
 This model is invariant under the supersymmetry transformations, so the
 bosonic and fermionic loops are mutually cancelled, the partition
 function is equal to unity, and only the diagrams of the type depicted in
 fig.~\ref{cactus} survive.

 The Schwinger--Dyson equations for the two--point correlator of bosonic
 matrices can be obtained from the equality
 \[
 \int\,\prod_{\alpha=1}^{2}\,dW^{\alpha}\,d\dw_{\alpha}\,
 \frac{1}{N^2}\sum_{ij}\frac{\partial}{\partial B^{\alpha}_{~ij}}
 \left[
 \e^{-N\tr\left(\dw D^{-1}W-
 \frac{1}{2}g\dw P^+W\dw P^+W-
 \frac{1}{2}g\dw P^-W\dw P^-W\right)}
 B^{\gamma}_{~ij}\right]=0
 \]
 and can be rewritten in the form
 \begin{eqnarray}\label{sdi1}
 &\left\langle\frac{1}{N}\tr\left[(\db D^{-1})_{\alpha}
 B^{\gamma}\right]\right\rangle-
 g\left\langle\frac{1}{N}\tr\left[(\db P^+)_{\alpha}
 B^{\gamma}\dw P^+W\right]\right\rangle&\nonumber\\&
 -g\left\langle\frac{1}{N}\tr\left[(\db P^-)_{\alpha}
 B^{\gamma}\dw P^-W\right]\right\rangle
 =\delta_{\alpha}^{~\gamma}.&
 \end{eqnarray}

 Similar to sec.~\ref{1mm}, we utilise the Ward identities to close
eq.~\rf{sdi1}:
\[
 \left\langle\delta_{\eps}\left(F^{\beta}
 \db_{\delta}B^{\gamma}\db_{\sigma}\right)\right\rangle=0.
\]
 Multiplying this equality by
 $(P^{\pm})_{~\alpha}^{\delta}(P^{\pm}
 )_{~\beta}^{\sigma}$ we get after some algebra the following
 equation
 \begin{equation}\label{wardi1}
 \left\langle\frac{1}{N}\tr\left[(\db P^{\pm})_{\alpha}
 B^{\gamma}\dw P^{\pm}W\right]\right\rangle=
 \left\langle\frac{1}{N}\tr\left[(\db P^{\pm})_{\alpha}
 B^{\gamma}\right]\frac{1}{N}\tr\left(\db P^{\pm}B\right)
 \right\rangle.
 \end{equation}
 Substituting it into eq.~\rf{sdi1}, one obtains in the large $N$ limit the
 closed equation for the $2\times 2$ matrix
 \begin{equation}\label{dcab}
 \cd_{\alpha}^{~\gamma}=\left\langle\frac{1}{N}\tr
 \db_{\alpha}B^{\gamma}\right\rangle.
 \end{equation}
 From \rf{sdi1} and \rf{wardi1} we find
 \begin{equation}\label{sdi2}
 D^{-1}\cd-
 g\sp\left(P^+\cd\right)P^+\cd
 -g\sp\left(P^-\cd\right)P^-\cd={\bf 1}.
 \end{equation}

 For the symmetry reasons, the dressed coupling constants entering
 \eq{sdi2} should be equal to each other:
 \begin{equation}\label{defg+-}
 g\sp\left(P^+\cd\right)=g\sp\left(P^-\cd\right)
 =\frac{g}{2}\sp\cd
 \equiv\G.
\end{equation}
Multiplying \eq{sdi2} by $D$ from the left, we obtain
 \begin{equation}\label{sdi3}
 \cd-
 \G D\cd
 =D.
 \end{equation}
 Solving eq.~\rf{sdi3} for $\cd$, we find
 \begin{eqnarray}\label{dc}
 \cd&=&\left({\bf 1}
 -\G D
 \right)^{-1}D\nonumber\\
 &=&\frac{1}
 {1-2\e^{\beta }\G+2\sinh\! 2\beta\, \G^2}
 \left(\begin{array}{cc}
 \e^{\beta }-2\sinh\! 2\beta\, \G&\e^{-\beta }\\ \e^{-\beta }&
 \e^{\beta }-2\sinh\! 2\beta\, \G\end{array}\right).
 \end{eqnarray}
 Substitution of the solution \rf{dc} into the definition of
 $\G$ (eq.~\rf{defg+-}) gives the equation from which it should be
 determined:
 \begin{equation}\label{cubic}
 2\sinh\! 2\beta\, \G^3-2\e^{\beta }\G^2+(2g\sinh 2\beta +1)\G-g\e^{\beta }
 =0.
 \end{equation}
 The third order equation \rf{cubic}, together with eq.~\rf{dc}, completely
 determines the two--point correlators of the model.  Which one to choose
 from the three solutions of \eq{cubic} is dictated by matching with
 perturbation theory.

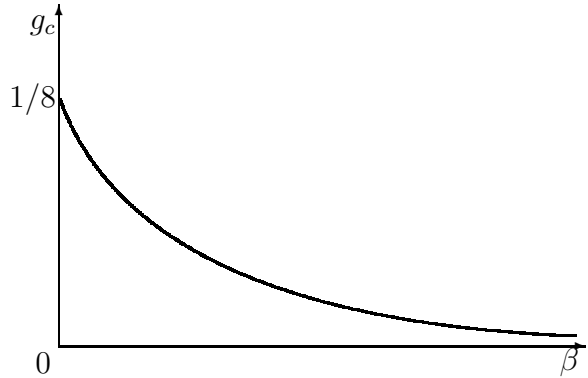
\begin{figure}
\unitlength .7mm
\linethickness{0.4pt}
\centering{
\begin{picture}(105.00,85.00)(20,40)
\put(30.00,50.00){\vector(1,0){100.00}}
\put(30.00,50.00){\vector(0,1){65.00}}
\thicklines
\bezier{508}(30.00,97.00)(45.00,55.33)(128.00,52.00)
\thinlines
\put(27.00,111.00){\makebox(0,0)[cc]{$g_c$}}
\put(27.00,47.00){\makebox(0,0)[cc]{$0$}}
\put(127.00,47.00){\makebox(0,0)[cc]{$\beta$}}
\put(25.00,97.00){\makebox(0,0)[cc]{$1/8$}}
\end{picture} }
\caption[cbp]{The phase diagram of the Ising model on branched polymer.}
\label{cr2mm}
\end{figure}

 The critical behaviour comes about when two roots of
 the equation \rf{cubic} collide. It happens when the following
 conditions hold simultaneously:
 \begin{eqnarray}\label{crit} 2\sinh\!
 2\beta\, \G_c^3-2\e^{\beta }\G_c^2+(2g_c\sinh 2\beta +1)\G_c-g_c\e^{\beta
 }&=&0,\nonumber\\ 6\sinh\! 2\beta\, \G_c^2-4\e^{\beta }\G_c+(2g_c\sinh
 2\beta +1)-g_c\e^{\beta }&=&0.
 \end{eqnarray}
The critical curve in the $\beta $--$g$ plane is depicted in fig.~\ref{cr2mm}.
 The critical coupling constant behaves in the low temperature limit, $\beta
 \rightarrow \infty $, as $g_c(\beta )\sim\frac{1}{4}\e^{-2\beta }$.
  In this limit the mixing between $W^1$ and $W^2$ is
 exponentially small and the above result reproduces the critical
 point of the solution \rf{f1g}, \rf{fng} of the one--matrix model. The
 two--point correlators always have the square root singularity along the
 line of the phase transition; this recovers, in particular, the
 known result that
 the Ising model on the branched polymer never becomes critical \cite{ising}.

\newsection{Discussion}
We have shown that the simplest non-trivial
supersymmetric matrix model produces abstract branched
polymers for the general potential~\rf{V}.
Certain aspects of these abstract branched polymers
have been found earlier by purely combinatorial methods, but the
matrix model method allows  a detailed study of the spectral properties.
We predict that the non-trivial phases of the abstract branched polymers,
corresponding the higher critical points with $m > 2$, also
exist in ``real'' lattice animals, i.e.\ branched polymers
embedded in regular $d$-dimensional lattices. Since these
branched polymers look increasing similar to ordinary polymers
for large $m$ (see \cite{multicr}) there might exist
a whole sequence of universality classes of lattice animals,
interpolating between ordinary polymers with internal (Hausdorff)
dimension 1 and ``conventional'' branched polymers with
internal (Hausdorff) dimension 2.

In field theory the use of matrix models has usually been associated
with Riemann surfaces. The large $N$ limit of Hermitian or complex matrix
models has a diagrammatic expansion as graphs of spherical topology and
an $1/N$ expansion is an expansion in diagrams with the topology
of higher genus surfaces. Adding Supersymmetry to
a purely bosonic or  purely fermionic theory makes it more rigid and
reduce the number of degrees of freedom. Here we observe this
in an extreme way: only the tiny subclass of graphs corresponding
to branched polymers survives for the model~\rf{z}, \rf{V}
when we look at observables like
$\tr 1/(z-\dbb)$.

It is an interesting question  whether or not supersymmetric matrix models
of the type discussed in this paper can be associated with super-Riemann
surfaces. As is already mentioned in the previous paragraph,
the potential \rf{V}  is separately
invariant under both~\rf{susy1} and \rf{susy2}.
This is the reason only  
cactus-like diagrams rather than planar graphs survive for the potential~\rf{V}
in the large $N$ limit.
We can reduce the symmetry by modifying the potential
in order for   a wider class of graphs to survive.
Such supersymmetric matrix models deserve further investigations.

\subsection*{Acknowledgements}

We are grateful to P.~Orland and G.~Semenoff for useful discussions.
This work was supported in part by the grant INTAS--94--840. The work of
K.~Z.\ was supported in part by RFFI grant 96--01-00344.
J.~A.\ acknowledges the support of the Professor Visitante Iberdrola Grant and
the hospitality at the University of Barcelona, where part of
this work were done.

\eop


\begin{thebibliography}{33}

\bibitem{disloc1}S. Iida, H.A. Weidenm\"{u}ller and J.A. Zuk,
Ann.\ Phys. 200 (1990) 219.
\bibitem{disloc2}J.A. Zuk, Phys.\ Rev. B45 (1992) 8952.
\bibitem{disloc3}A. M\"{u}ller-Groeling, Phys.\ Rev. B47 (1993) 6480.
\bibitem{disloc4}A. Altland, S. Iida, A. M\"{u}ller-Groeling,
H.A. Weidenm\"{u}ller, Ann.\ Phys. 219 (1992) 148.
\bibitem{qwires}A.D. Mirlin, A. M\"{u}ller-Groeling, and M.R. Zirnbauer,
Ann.\ Phys. 236 (1994) 325.
\bibitem{qhall}H.A. Weidenm\"{u}ller and M.R. Zirnbauer, Nucl.\ Phys. B305
(1988) 339.
\bibitem{wigner}E.P. Wigner, Ann.\ Math. 53 (1951) 36; Proc.\ Cambridge
Philos. Soc. 47 (1951) 790.
\bibitem{dyson}F.J. Dyson, J. Math.\ Phys. 3 (1962) 140.
\bibitem{mehta}M.L. Mehta, {\it Random Matrices}, 2nd edition
(Academic Press, N.Y. 1991).
\bibitem{nuclear}O. Bohigas and H.A. Weidenm\"{u}ller,
Ann.\ Rev.\ Nucl.\ Part.\ Sci. 38 (1988) 421.
\bibitem{chiral}J.J.M.  Verbaarschot and I. Zahed, Phys.\ Rev.\ Lett.
70 (1993) 3852.\\
E.V. Shuryak and J.J.M.  Verbaarschot, Nucl.\ Phys. A560 (1993) 306.\\
J. Jurkiewicz, M.A. Nowak and I. Zahed, hep-ph/9603308.
\bibitem{chaos}M. Gutzwiller, {\it Chaos in classical and quantum
mechanics}, Springer--Verlag, 1990.
\bibitem{gravity}F.  David,
{\it Simplicial quantum gravity and random lattices.}
Les Houches Sum.\ Sch.1992:679-750 (QC178:H6:1992),
hep-th/9303127.\\
P. Ginsparg and G. Moore, {\it Lectures on 2-d gravity and 2-d string theory.}
TASI 92:277-470 (QCD161:T45:1992), hep-th/9304011.\\
P. Di Francesco, P. Ginsparg and J. Zinn-Justin, Phys.\ Rep. 254 (1995) 1.
\bibitem{bipz}
E. Brezin, C. Itzykson, G. Parisi and Z.B. Zuber, Commun.\ Math.\ Phys. 59
(1978) 35.
\bibitem{ackm}
J. Ambj\o rn, L. Chekhov, C.F Kristjansen and Yu.\ Makeenko,
Nucl.\ Phys. B404 (1993) 127.
\bibitem{ajm}
J. Ambj\o rn, J. Jurkiewicz and Yu.M. Makeenko,
 Phys.\ Lett. B251 (1990) 517.
\bibitem{zee}E. Brezin and A. Zee, Nucl.\ Phys. B402 (1993) 613;
Phys.\ Rev. E49 (1994) 2588.
\bibitem{efetov}K.B. Efetov, Adv.\ Phys. 32 (1983) 53.
\bibitem{ver}J. Verbaarschot, H.A. Weidenm\"{u}ller and M.R. Zirnbauer,
Phys.\ Rep. 129 (1985) 367.
\bibitem{zuk}J.A. Zuk,
{\it Introduction to the supersymmetry method for the Gaussian random
matrix ensembles}, cond-mat/9412060.
\bibitem{enzo}
E. Marinari and G. Parisi,
Phys. Lett. B240 (1990) 375.
\bibitem{luis} L. Alvarez-Gaume and  J.L. Manes,
Mod.\ Phys.\ Lett. A6 (1991) 2039.
\bibitem{MP95}
Y. Makeenko and Hla Win Pe,
{\it Supersymmetric matrix models and the meander problem},
hep-th/9601139.

\bibitem{cardy}J.L. Cardy, {\it Phase Transitions and Critical
Phenomena}, Vol 11, ed. C. Domb and J.L. Lebowitz (London, Academic Press).
\bibitem{parisi}G. Parisi and N. Sourlas, Phys.\ Rev.\ Lett. 46 (1981) 871. \\
J.D. Miller and K. De'Bell, J. Physique, I 3 (1993) 1717.

\bibitem{branched}
M. Giri, M.J. Stephen and G.S. Grest, Phys.\ Rev. B16 (1977) 4971.\\
F.Y. Wu, Rev.\ Mod.\ Phys. 54 (1982) 235.\\
H.K. Janssen and A. Lyssy, Phys.\ Rev. E50 (1994) 3784.\\
M. Henkel and F. Seno, cond-mat/9601105.

\bibitem{onboson}S. Nishigaki, Mod.\ Phys.\ Lett. A9 (1991) 631.\\
S. Nishigaki and T. Yoneya, Nucl.\ Phys. B348 (1991) 787.\\
P. DiVecchia, M. Kato and N. Ohta, Nucl.\ Phys. B357 (1991)

\bibitem{onfermion}
G.W. Semenoff and R.J. Szabo,
{\it Polymer statistics and fermionic vector models.},
hep-th/9602007.

\bibitem{polymer}
J.~Ambj{\o}rn, B.~Durhuus, J.~Fr{\"o}hlich, and P.~Orland,
{Nucl.~Phys.} {B270 [FS16]} (1986) 457.

\bibitem{multicr}
J.~Ambj{\o}rn,
B.~Durhuus, and T.~J{\'o}nsson, {Phys.~Lett.} {B244} (1990) 403.

\bibitem{ising}
J.~Ambj{\o}rn, B.~Durhuus, T.~J{\'o}nsson, and G.~Thorleifsson,
{Nucl.~Phys.} {B398} (1993) 568.


\end{thebibliography}
\end{document}